\documentclass{aa}  
\usepackage{graphicx}
\usepackage{txfonts}
\usepackage{hyperref}

\newcommand{\g}{$\gamma$}
\usepackage{lineno}

\begin{document}

   \title{Search for AGN counterparts of unidentified \textit{Fermi}-LAT sources with optical polarimetry}

   \subtitle{Demonstration of the technique}

   \author{N.~Mandarakas\inst{1}\fnmsep\thanks{\email{nmandarakas@physics.uoc.gr}}
          \and
          D.~Blinov\inst{1,2,3}
          \and
          I.~Liodakis\inst{4}
          \and
          K.~Kouroumpatzakis\inst{1,2}
          \and
          A.~Zezas\inst{1,2}
          \and
          G.~V.~Panopoulou\inst{5}
          \and
          I.~Myserlis\inst{6}
          \and
          E.~Angelakis\inst{6}
          \and
          T.~Hovatta\inst{7}
          \and
          S.~Kiehlmann\inst{5}
          \and
          K.~Kokolakis\inst{1,8}
          \and
          E.~Paleologou\inst{1,2}
          \and
          A.~Pouliasi\inst{1}
          \and
          R.~Skalidis\inst{1,2}
          \and
          V.~Pavlidou\inst{1,2}
          }

   \institute{Department of Physics and Institute for Theoretical and Computational Physics (ITCP), University of Crete, 71003, Heraklion, Greece
         \and
             Foundation for Research and Technology - Hellas, IESL, Voutes, 7110 Heraklion, Greece
         \and
             Astronomical Institute, St. Petersburg State University, Universitetsky pr. 28, Petrodvoretz, 198504 St. Petersburg, Russia
         \and
             KIPAC, Stanford University, 452 Lomita Mall, Stanford, CA 94305, USA
        \and
             Owens Valley Radio Observatory, California Institute of Technology, Pasadena, CA 91125, USA
         \and
             Max-Planck-Institut f\"{u}r Radioastronomie, Auf dem H\"{u}gel 69, 53121 Bonn, Germany
         \and
         Tuorla Observatory, Department of Physics and Astronomy, University of Turku, Finland
         \and
            Geodesy \& Geomatics Engineering Lab, Technical University of Crete, GR-73100, Chania, Greece
             }

   \date{Received October 18, 2018; accepted January 15, 2019}

% \abstract{}{}{}{}{} 
% 5 {} token are mandatory
 
  \abstract
   {The third \textit{Fermi}-LAT catalog (3FGL) presented the data of the first four years of observations from the \textit{Fermi} Gamma-ray Space Telescope mission. There are 3034 sources, 1010 of which still remain unidentified. Identifying and classifying \g-ray emitters is of high significance with regard to studying high-energy astrophysics.}
   {We  demonstrate that optical polarimetry can be an advantageous and practical tool in the hunt for counterparts of the unidentified \g-ray sources (UGSs).}
   {Using data from the RoboPol project, we validated that a significant fraction of active galactic nuclei (AGN) associated with 3FGL sources can be identified due to their high optical polarization exceeding that of the field stars. We performed an optical polarimetric survey within $3\sigma$ uncertainties of four unidentified 3FGL sources.}
   {We discovered a previously unknown extragalactic object within the positional uncertainty of 3FGL~J0221.2+2518. We obtained its spectrum and measured a redshift of $z=0.0609\pm0.0004$. Using these measurements and archival data we demonstrate that this source is a candidate counterpart for 3FGL~J0221.2+2518 and most probably is a composite object: a star-forming galaxy accompanied by AGN.}
   {We conclude that polarimetry can be a powerful asset in the search for AGN candidate counterparts for unidentified \textit{Fermi} sources. Future extensive polarimetric surveys at high galactic latitudes (e.g., PASIPHAE) will allow the association of a significant fraction of currently unidentified \g-ray sources.}


   \maketitle
%
%-------------------------------------------------------------------

\section{Introduction}

Since the launch of the \textit{Fermi} spacecraft on 11 June 2008, a vast amount of data has been 
collected on \g-ray sources. The entire set of point sources detected during the first four years of 
observations is presented in the 3FGL catalog \citep{Acero2015}. Among the 3034 sources in 3FGL, 
about one-third (1010) are still unassociated with low-energy counterparts, while AGN account for 
$\sim 85\%$ of the associations and identifications ($\sim57\%$ of the entire sample of 3034 
sources). In order to classify and associate \g-ray sources, various techniques have been used, as 
summarized below. 

{\bf Machine learning.} In \cite{Doert&Errando2014}, machine learning algorithms were used to find 
objects with AGN-like properties in the unassociated sources of the 2FGL catalog. Machine training 
was conducted using 70\% of the known AGN in the catalog, while the remaining 30\% were used for 
testing. Results showed that the algorithm is expected to recognize 80\% of the AGN present in the 
unassociated sample, with a false-association rate of 11\%. This technique provided a total of 231 
new AGN candidates among the 576 unassociated sources that were studied.

\cite{Chiaro2016} and \cite{Salvetti2017} used the \g-ray variability properties of unassociated 
sources and neural networks in order to classify these sources. They demonstrated that the 
percentage of sources of uncertain type in 3FGL can be decreased from 52\% to 10\% with the use of 
their method. Similar classification of UGSs can be useful for optimization of surveys dedicated for 
their identification.

{\bf VLBI observations.} \cite{Kovalev2009} proposed using very long baseline interferometry (VLBI) 
for identification of \g-ray sources. The author cross-correlated positions of 205 \g-ray loud 
sources observed by \textit{Fermi}-LAT with VLBI coordinates of a large sample of extragalactic 
sources. He was able to confirm the findings of LAT and suggest six new identifications.

{\bf Multiwavelength studies.} \cite{Acero2013} studied \textit{Fermi}-LAT sources that had also 
been observed by the Swift satellite with its X-ray telescope (XRT). Swift XRT allowed precise 
localization at the level of a few arcseconds, with the detected sources being then observed in the 
radio, IR, or optical. Seven high-latitude sources were investigated, four of which were found to be 
AGN candidates and  one a pulsar candidate. The authors speculated that the two remaining objects 
may belong to a new category subclass or point to a new type of \g-ray emitter.

{\bf Radio spectra.} In their search for pulsars in the 3FGL sample, \cite{Frail2016} examined radio 
spectra of unidentified sources within the 95\% confidence error ellipses, using existing catalogs. 
Compact objects that are bright in MHz frequencies but faint in GHz frequencies were categorized as 
pulsar candidates.

{\bf Radio observations.} \cite{Barr2013} conducted radio observations of 289 unassociated sources 
from the 1FGL catalog using the Effelsberg radio telescope in a search for pulsars. Objects studied 
were located in the center of their 95\% confidence ellipses. Using this method, one millisecond 
pulsar was discovered.

 \cite{Schinzel2017}, using the Australia Telescope Compact Array and Very Large Array in the range 
of 4.0-10.0 GHz, performed a survey of all unidentified \textit{Fermi} sources in the 3FGL catalog, 
in their search for radio counterparts. They found 2097 candidates, with several fields containing 
multiple compact radio sources, while others  did not contain any  above 2 mJy. For several of these 
targets they performed follow-up observations with VLBI, which provided 142 new AGN associations,  
alternative associations for 7 objects, improved positions for 144 known associations, as well as 36 
extended radio sources. Among the fields studied was 3FGL~J0221.2+2518, which is the field of 
interest of this paper. They propose two possible radio counterparts lying within this field. We 
discuss the possibility of these associations with the \textit{Fermi} source in Sect. 
\ref{sec:candidates}.

{\bf Figure of merit (FoM).} \cite{SowardsEmmerd2003} used a FoM approach to quantify the 
probability that an unassociated source is a blazar. 
To form this FoM, basic characteristics of blazars are taken into account: radio and X-ray 
properties as well as source position. Based on this approach, the authors evaluated associations of 
\g-ray and radio sources and presented $\sim$20 new identifications.

These methods for the identification of \textit{Fermi} sources make use of various characteristics 
of \g-ray emitters. Optical polarization is a frequent trait of \g-ray sources that has yet to be 
exploited in the search for candidate counterparts of yet-unassociated sources. 

Blazars are a subclass of AGN with powerful relativistic jets oriented towards our line of sight, 
which causes strong relativistic boosting of their synchrotron radiation \citep{Blandford1979}. Due 
to the synchrotron nature of their optical emission, blazars are often highly polarized in the 
optical band \citep{Angel1980,Angelakis2016}. Since blazars constitute the majority of \g-ray 
sources, $\sim 85\%$ of the identified or associated sources and $\sim57\%$ of the entire 3FGL 
catalogue \citep{Acero2015}, it is extremely important to be able to distinguish them from other 
star-like sources in UGS fields. In the next section we investigate the potential of optical 
polarimetry as a new method for the identification of blazars responsible for UGSs.

The values of the cosmological parameters adopted throughout this work are $H_0 = 67.8$ km s$^{-1}$ 
Mpc$^{-1}$, $\Omega_m = 0.308$, and $\Omega_\Lambda = 1 - \Omega_m$ \citep{Ade2016}.

\section{Optical polarimetry as a tool for identification of UGSs}
\subsection{Blazar detection efficiency} \label{subsec:effic}
While blazars are typically moderately to highly polarized in the optical, they are not  the only 
type of source that can appear polarized in the optical band. We must therefore take into account 
all processes that produce polarization in the optical, characterize their properties, and finally 
select the characteristics that isolate blazars from other types of polarized sources. In any given 
line of sight, light passing through the galactic interstellar medium (ISM) becomes linearly 
polarized due to dichroic extinction from dust grains that are aligned with the interstellar 
magnetic field \citep[for a recent review see][]{Andersson2015}. The linear polarization fraction 
induced by the ISM is typically at a level of a few percent. This can be enough to hinder the 
identification of a blazar within a typical field. Additionally, the intrinsic fractional 
polarization of blazars is known to be variable, which can also make them indistinguishable if 
observed only once at their low-polarization state.

In order to evaluate the efficiency of our method, we developed a Monte Carlo (MC) simulation that 
allowed us to investigate whether a blazar would be significantly more polarized than foreground 
stars. To account for the fact that different parts of the sky exhibit different average 
interstellar polarization and the polarimetric properties in a single region vary between stars in 
the same region, we relied on the detailed, high-accuracy optopolarimetric mapping of the well-known 
Polaris Flare cloud using the RoboPol instrument \citep{Panopoulou2015}, and rescaled its 
polarization properties to different average polarization values that may be applicable at 
different Galactic latitudes.  

\begin{figure}
   \centering
   \includegraphics[width=0.44\textwidth]{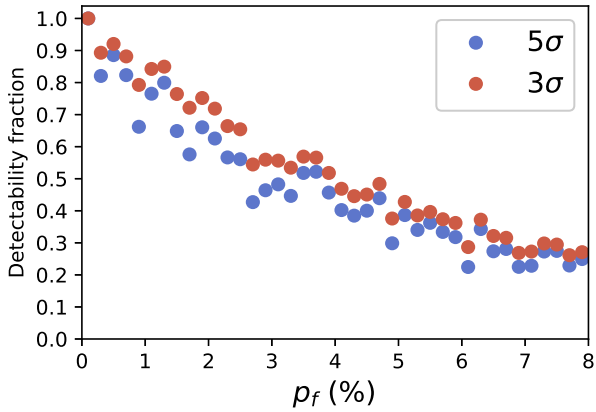}
      \caption{Percentage of blazars that will stand out $3\sigma$ and $5\sigma$ from the average 
polarization of the field $p_f$.}
         \label{frac_ips}
\end{figure}

First we estimated how much interstellar polarization varies from star to star in an area of the sky 
that is typical for UGSs position uncertainty. To this end, we found that this area for sources in 
the 3FGL catalog is 0.0456 deg$^2$, which corresponds to a circle with a radius of 0.12 deg. We 
placed this circle in random positions in the Polaris Flare cloud region and measured the standard 
deviation of fractional polarization, $\sigma_p$, of stars within it, on the condition that there 
are five or more stars with measured polarization by \citet{Panopoulou2015} within the selected 
area. We repeated the process until we obtained a set of 5000 $\sigma_p$ values from which we 
calculated the standard deviation $\sigma_{\sigma_p}$ and the mean $M_{\sigma_p}$.
Then assuming that at any position on the sky  interstellar polarization has the same variance as in 
the Polaris Flare region, i.e., following the normal distribution 
$\mathcal{N}(M_{\sigma_p},\sigma_{\sigma_p})$, we performed the MC simulation as follows:
\begin{enumerate}
\item We generated values representing average field polarization in UGSs fields, $p_f$, in the 
range [0\%,8\%],  
with a step of 0.2\%. For each  simulated average field polarization we assigned a random 
$\sigma_f$ taken from $\mathcal{N}(M_{\sigma_p},\sigma_{\sigma_p})$ found before.
\item For every value of the average field interstellar polarization (ISP) we drew a random blazar 
and its intrinsic average polarization $p_0$ and modulation index $m_p$ from 
the list of 62 \g-loud blazars presented in \citet{Angelakis2016}. This sample is a $\gamma$-ray 
photon-flux limited subsample of 2FGL blazars. It was selected using strict and unbiased criteria 
making it a representative sample of the parent population of \g-loud blazars.
\item In order to account for their variability properties, for each blazar selected in step 2, we 
drew a random value for its polarization degree ($p_{gen}$) from a  Beta distribution  
\citep{Blinov2016}:
\begin{equation}
\alpha(p_0,m_p) = \left ( \frac{1-p_0}{p_0 m_p^2} - 1 \right ) p_0,
\end{equation}
\begin{equation}
\beta(p_0,m_p) = \left ( \frac{1-p_0}{p_0 m_p^2} - 1 \right ) (1 - p_0).
\end{equation}
\item We considered a blazar to be significantly more polarized than the field stars (i.e., 
detectable) if $p_{gen} > p_f + \mathrm{SL} \times \sigma_f$, where the significance level (SL) is 
the number of standard deviations.
\end{enumerate}
Repeating the simulation $10^3$ times for SL = 3 and 5, we found the expected 
fraction of the fields where the UGS could be detected (in the case where the UGS is associated with 
a blazar) using optical polarization measurements. The results of the simulation are shown in 
Fig.~\ref{frac_ips}. It follows from this plot that for high galactic latitudes 
($|\rm{b}|>10$\degr), where the average field polarization is expected to be $\le 1\%$, $\sim$80\% 
of the blazars in UGSs will be 3$\sigma$ more polarized than the background ISP.

\subsection{Expected number of detections}
In order to estimate the number of sources among UGSs that can be detected in a polarimetric survey, 
we performed the following MC simulation. For each UGS we found the reddening $E(B-V)$ from 
\cite{Schlafly2011}. Then we estimated the maximum possible ISP value for each field following the 
relation by \cite{Hiltner1956}:
\begin{equation} \label{eq:P_EBV}
P_{\rm max} \le 9 E(B-V) (\%/mag).
\end{equation}
After assigning random optical polarization $P_r$ for each source following the procedure describe 
dabove, we considered an UGS to be suitable for detection when $P_r \ge P_{\rm max}$. Repeating this 
simulation $10^4$ times we found that if $\sim85\%$ of unidentified sources are blazars (as it is 
for identified sources in 3FGL), then $526\pm9$ could be detected using optical polarimetry.

We note that equation~\ref{eq:P_EBV} significantly overestimates ISP for high extinction regions 
because high extinction values are reached in the case of multiple foreground dust screens, while it 
is unlikely that the magnetic field is perfectly aligned within these regions with respect to each 
other. Therefore, above some level of $E(B-V)$ the increase of ISP halts due to depolarization 
caused by the diverse magnetic field directions in different polarizing screens. Moreover, the 
recalibration of extinction maps of \cite{Schlegel1998} provided by \cite{Schlafly2011} may be 
inaccurate outside of the Sloan Digital Sky Survey footprint \citep{Schlafly2014}. These two factors 
lead us to  repeat the estimation of the number of detectable sources using a different 
approach.

\begin{figure}
 \centering
 \includegraphics[width=0.43\textwidth]{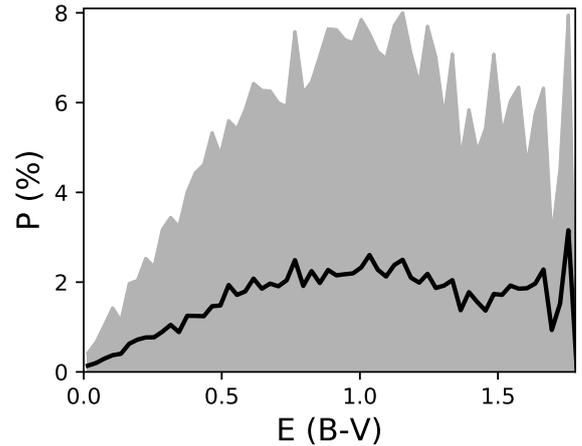}
 \caption{Dependence of stellar polarization from \cite{Heiles2000} with the corresponding reddening 
 from \cite{Schlafly2014}. The black line represents the mean polarization of stars within 0.03 mag 
bin of E(B-V). The gray area shows 3 standard deviations from the mean.}
 \label{P_EBV}
\end{figure}

Using 5590 stars with high signal-to-noise ratio  measurements ($P/\sigma P > 3$) from 
\cite{Heiles2000} and the dust reddening map by \cite{Schlafly2014} we found the dependence of $P$ 
on $E(B-V)$. For each star with a polarization measurement we found the corresponding reddening 
value, then we split the entire range of reddening  into bins of 0.03 mag and calculated the mean 
$p_{redd}$ and the standard deviation $\sigma_{redd}$ of polarization for the stars within each bin. 
The obtained dependence is shown in Fig.~\ref{P_EBV}.
Then for each UGS we found $E(B-V)$ from \cite{Schlafly2014}\footnote{In the case when it was 
outside the footprint of \cite{Schlafly2014}, we used $E(B-V)$ from \cite{Schlafly2011}.} and 
assigned a corresponding $p_{redd}$ and $\sigma_{redd}$ using the dependence in Fig.~\ref{P_EBV}. 
For UGSs with $E(B-V)>1.75$, which is  outside  the range covered by \cite{Heiles2000}, stars were 
assigned $p_{redd} = 100\%$. After that we repeated  step 4 from Sect.~\ref{subsec:effic} and 
assigned each source a random polarization $P_r$ following the beta distribution and parameters of 
real blazars from \cite{Angelakis2016}. Then we repeated the last step $10^4$ times computing the 
fraction of sources where $P_r > p_{redd} + 3 \times \sigma_{redd}$. We found that in the case where 
85\% of currently unidentified 3FGL sources are blazars, we expect to be able to detect $544 \pm 10$ 
of them in an optical polarization survey, which is consistent with the previous  rougher estimate.

It is worth noting that our simulations take into account only polarization degree while the 
direction of polarization plane is omitted. A more accurate approach must take into account the 
vector nature of linear polarization. Observed polarization of a blazar is a vector sum of its 
intrinsic and the interstellar polarizations. For this reason even blazars with intrinsic 
polarization values lower than the average interstellar field polarization can be detected if the 
polarization angles of the two significantly differ. Therefore, the efficiency of the technique may 
in fact be even higher than the estimate presented above.

\subsection{Survey strategy}
Before proceeding to put the method to the test, we established a ranking parameter to quantify the 
possibility of detecting a blazar in a \textit{Fermi}-LAT region. In other words, we characterized 
fields by suitability for follow-up observation. We define the ranking 
parameter as $R=(Var \times Flx \times 10^6)/Pos$. The variables affecting this parameter are the 
following:
\begin{itemize}
\item \textit{Variability (Var)}: Sources from 3FGL are flagged with a \textit{Var} index, which 
denotes the probability of a source to be variable \citep{Acero2015}. High values of \textit{Var} 
denote a higher probability that the source is an AGN.
\item \textit{\g-ray photon flux (Flx)}: Given our current understanding of the relation between 
optical and \g-rays, a high photon flux in the range 100~MeV -- 100~GeV would suggest a high optical 
flux density if the source responsible for the \g-ray emission is an AGN 
\citep[e.g.,][]{Cohen2014,Liodakis2018}.
\item \textit{Positional error (Pos)}: The positional error determined as the area of an ellipse 
encircling the 95\% confidence region of \textit{Fermi} sources locations. It is given in $deg^2$ 
and is calculated as $\pi \times a \times b$, where $a$ and $b$ are semi-major and semi-minor axes 
of the positional error ellipse provided in 3FGL. The larger the positional error, the more 
difficult it is to study the area.
\end{itemize}
The ranking parameter ensures that the target field under investigation contains a variable, \g-ray 
bright source within a relatively small region of the sky. 

We observed fields of four UGSs with various ranking parameters. These sources were selected 
randomly among all visible UGSs at the moment of observations. Their ranking parameters and the 
sequential positions in the list of 1010 unidentified \textit{Fermi} sources sorted by the rank are 
presented in Table~\ref{tab:rank}.

\begin{table}
\caption{Data information and ranking parameters of the four observed sources along with their 
sequential position in the list sorted by ranking. Flx is measured in $\mathrm{10^{-8} ph}$ 
$\mathrm{{cm}^{-2} s^{-1}}$ and Pos in $\mathrm{deg^{2}}$.}
\label{tab:rank}
\centering
\begin{tabular}{c c  c c c c}
\hline
3FGL id & Flx & Var & Pos &Rank.Par. & Position \\
\hline
   J1848.6+3232 & 2.84 & 193 & 0.01 &551 & 31  \\
   J0419.1+6636 & 2.2 & 49 & 0.008 &130 & 106  \\
   J0336.1+7500 & 1.06 & 34 & 0.007 &52  & 219  \\
   J0221.2+2518 & 0.45 & 38 & 0.024 &7   & 660  \\
\hline
\end{tabular}
\end{table}

\section{Observations and data reduction} \label{sec:obs}

Polarimetric data of the targets were obtained using the RoboPol\footnote{\url{http://robopol.org/}} 
polarimeter attached to the 1.3m. telescope at the Skinakas observatory (35.2120\degr N, 
24.8982\degr E) located in Crete, Greece. RoboPol contains a combination of two Wollaston prisms and 
two half-wave plates simultaneously splitting  incoming light in four different polarization 
directions, which are then projected as a four-point image for each source on the CCD. The only 
moving part is its filter wheel, which is equipped with B, V, R, I Johnson-Cousins filters. The 
particular design of the RoboPol polarimeter allows the measurement of the Stokes parameters with a 
single exposure, thereby minimizing systematic and statistical errors. The instrument is optimized 
for measurements of a source at the center of its $13\arcmin \times 13\arcmin$ field of view by  a 
mask in the telescope focal plane. The mask has a cross-shaped aperture in the center and is 
designed to block unwanted photons from the nearby area of the central source, as well as nearby 
sources from overlapping with it. The background noise surrounding the spots is reduced by a factor 
of 4 compared to field sources, allowing more precise and reliable measurements. RoboPol was 
primarily designed to monitor the optical linear polarization of blazars, with the first 
observations taking place in June 2013 \citep{Pavlidou2014}.

The operation of the instrument and data reduction is based on an automated pipeline described in 
detail by \cite{King2014}. Although the pipeline processes the entire RoboPol field of view, there 
are certain issues that need to be taken into account when performing and analyzing field 
measurements. The issues affecting our measurements are briefly discussed below.

\begin{figure}
   \centering
   \includegraphics[width=0.50\textwidth]{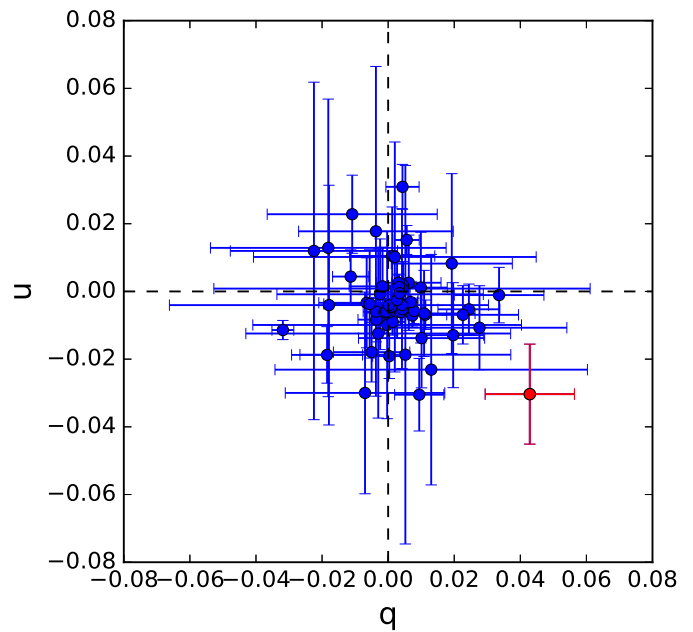}
      \caption{Distribution of q and u Stokes parameters of field sources. UGSC is shown by the red 
symbol.}
         \label{QU}
\end{figure}

$\bullet$ \textit{Large scale optical aberrations}: Aberrations caused by the optical system are 
corrected by the instrument model described in \cite{King2014}, and improved by 
\cite{Panopoulou2015}. In the latter paper there is also an estimate of the residual uncertainty 
after the instrumental model correction.

$\bullet$ \textit{Proximity of two sources}: Since RoboPol produces a four-point image for each 
source, it is common for one or more of these points to overlap with a point from a nearby source. 
Such sources are excluded from the analysis on condition that a spot exists within 3$\times$FWHM of 
another source's spot.

$\bullet$ \textit{Proximity to the CCD edges}: Sources close to the CCD edges are very likely to 
suffer from partial photon losses; i.e., one or more of the four spots are not projected on the CCD 
image. Consequently, sources falling 100 pixels or less from the edges are rejected from the 
analysis.

$\bullet$ \textit{Aperture optimization}: Stokes parameters $q{=}Q/I$ and $u{=}U/I$ are calculated 
through aperture photometry in each of the four spots of the same source. A number of conditions may 
affect the PSF of the spots (e.g., weather, seeing, optical system); therefore,  it is necessary to 
employ different photometry parameters for each of the spots. We account for this using an aperture 
optimization algorithm \citep{Panopoulou2015}. 

$\bullet$ \textit{Dust specks}: Telescope and RoboPol optics could be contaminated with dust 
resulting in specks on the produced CCD image. Objects falling on the dust specks are removed from 
the analysis.
   
For a more detailed description of errors in field measurements and corresponding solutions, refer 
to \cite{Panopoulou2015}.

We performed measurements of the sources within $3\sigma$ positional uncertainties of the four UGSs 
listed in Table~\ref{tab:rank} using the RoboPol instrument in August -- October 2017. Observations 
were conducted in the R band. For each field we obtained 3$\times$190 sec exposures at five 
positions of the telescope, separated by 1.2 arcmin, and having a square shape with one pointing in 
the center.

\section{Results}  
  \begin{figure}
   \centering
   \includegraphics[width=0.45\textwidth]{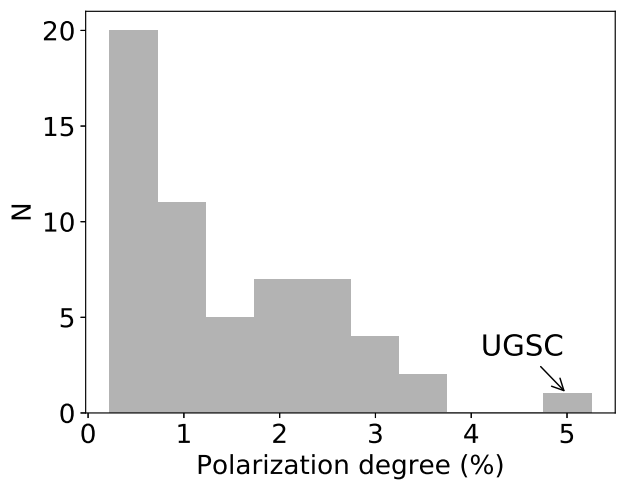}
      \caption{Distribution of the polarization values for the 57 sources in the 3FGL~J0221.2+2518 
field.}
         \label{p_hist}
   \end{figure}

For the first three UGS fields, despite their high ranking parameters, we do not find any source 
that has significantly higher optical polarization than the average value for the field. However, in 
the case of 3FGL~J0221.2+2518 we detect an outlier. Here we focus only on this UGS.

Analysis of the field provided us with reliable polarization measurements for 57 sources in the 
field of interest. Although there are many more objects within 3FGL~J0221.2+2518 field, sources 
fainter than 18th magnitude were not taken into account due to their high measurement uncertainties. 
In addition, a number of moderately bright sources were excluded from the analysis or could not be 
resolved due to issues discussed in Sect.~\ref{sec:obs}. After processing the results of the 
analysis, we observed the objects with the highest polarization in the mask of RoboPol to acquire 
more accurate measurements.

  \begin{figure}
   \centering
   \includegraphics[width=0.50\textwidth]{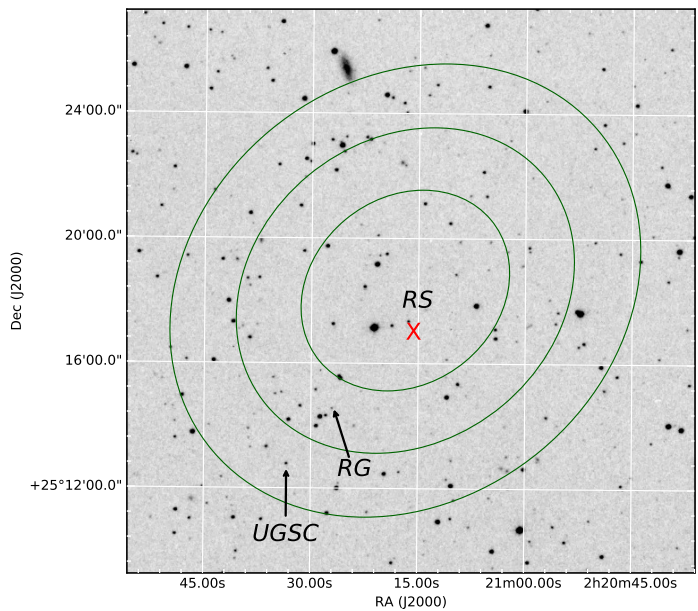}
      \caption{DSS image of 3FGL~J0221.2+2518 field and the position of UGSC and other candidates 
for the field as discussed in Sect.~\ref{sec:candidates}. \textit{RS} and \textit{RG} are the two 
candidates proposed by \cite{Schinzel2017} and stand for radio source and radio galaxy, 
respectively. The ellipses represent $1\sigma$, $2\sigma$, and $3\sigma$ uncertainties of the \g-ray 
source position.}
         \label{Field}
   \end{figure}

The values of $q$ and $u$ are represented in Fig.~\ref{QU} and the corresponding distribution of 
fractional polarization values is shown in Fig.~\ref{p_hist}. There is a source with polarization 
degree 5.2$\pm$1.3\% clearly deviating from the median polarization degree across the field 
($p_{av}{=}0.91{\pm}0.07\%$) that qualifies as a \g-ray emitting candidate. The Unidentified 
Gamma-ray Source Candidate (UGSC) is located at RA=02h21m33.3s, Dec=+25\degr12\arcmin47.2\arcsec 
(J2000), and its position with respect to the \textit{Fermi} source is presented in 
Fig.~\ref{Field}. It is listed in SDSS as J022133.31+251247.3 with $r=17.59$ mag. This source is not 
presented in any known AGN catalogue. The polarization degree of UGSC (albeit moderate compared to 
bonafide blazars)  indicates synchrotron emission as its origin; however, other mechanisms are 
possible. For instance, the polarization degree up to 4\% can be reached due to the scattering in 
circumstellar discs of Be/X-ray binaries \citep{Halonen2013} or pre-main sequence stars 
\citep{Oudmaijer2001}. Therefore, we collected additional archival data and performed supplementary 
optical spectroscopic observations in order to validate its AGN nature, as presented in the 
following section.

\section{Evidence of AGN nature of the UGSC} \label{sec:evidence}
\subsection{Optical spectrum and redshift}

\begin{figure}
   \centering
   \includegraphics[width=0.50\textwidth]{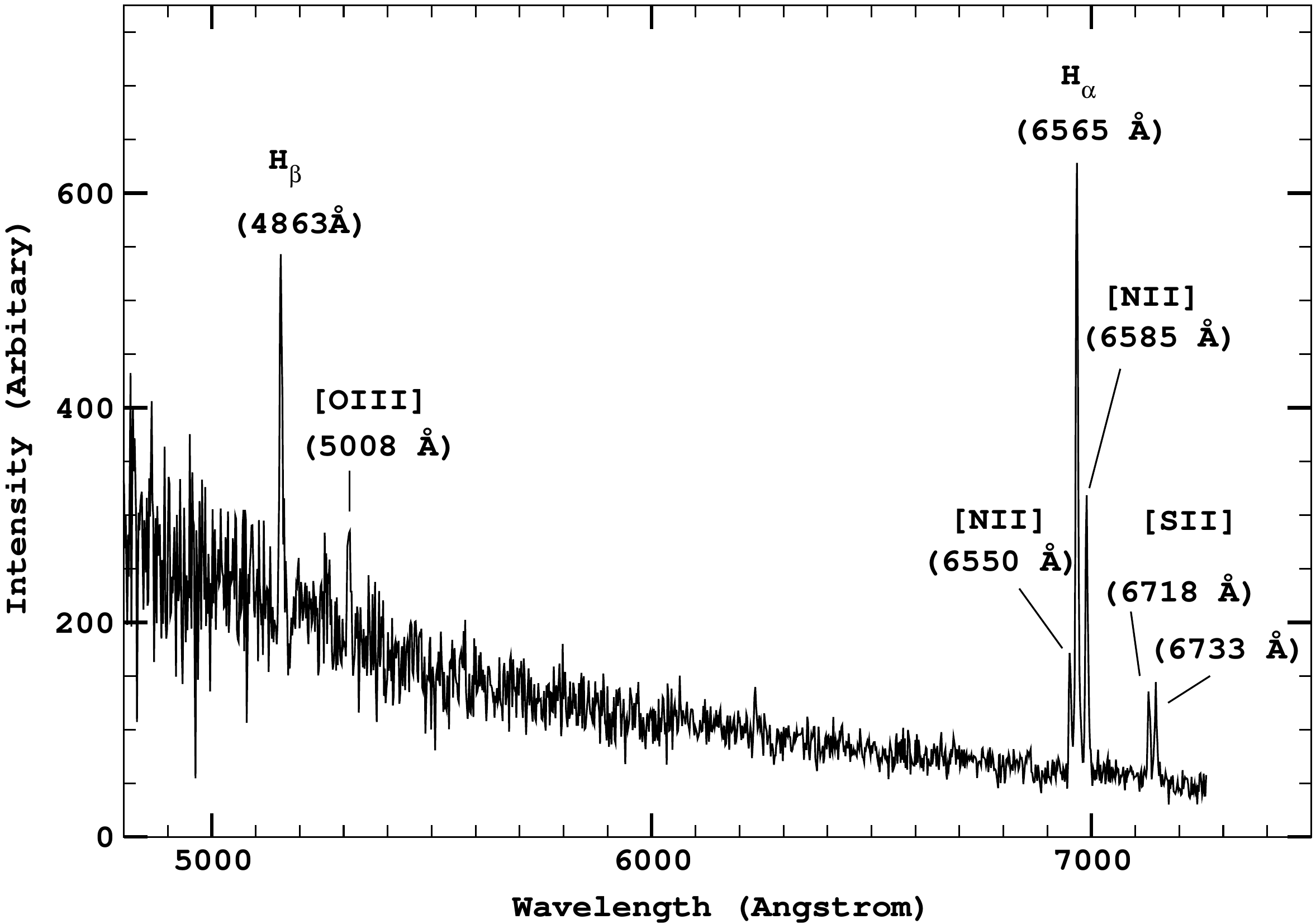}
      \caption{Optical spectrum of UGSC. Presented are the lines that were identified along with 
their rest frame wavelength.}
         \label{Spectrum}
   \end{figure}

Even though the UGSC is in the SDSS catalogue, it does not have an available spectrum. Therefore, 
we obtained a spectrum of this source, using the 1.3~m telescope of the Skinakas observatory. The 
spectrograph is equipped with an ANDOR DZ436 CCD camera with 2048$\times$2048 pixels and a 651 
lines/mm grating, giving a nominal dispersion of $\sim$1.85 {\AA}/pixel. The total exposure time was 
4500 sec divided in three exposures. The spectrum was processed using the standard IRAF (version 
2.16.1) CCD reduction, optimal extraction, and calibration. The spectrum is shown in 
Fig.~\ref{Spectrum}. It is not flux-calibrated since it is not necessary for performing line 
intensity ratio calculations. Different distortions can be caused in the intensity of the spectrum 
in different wavelengths. In our case, emission lines are close, thus their relative intensity 
weakly depends on the wavelength.

\begin{figure*}
   \centering
   \includegraphics[width=0.79\textwidth]{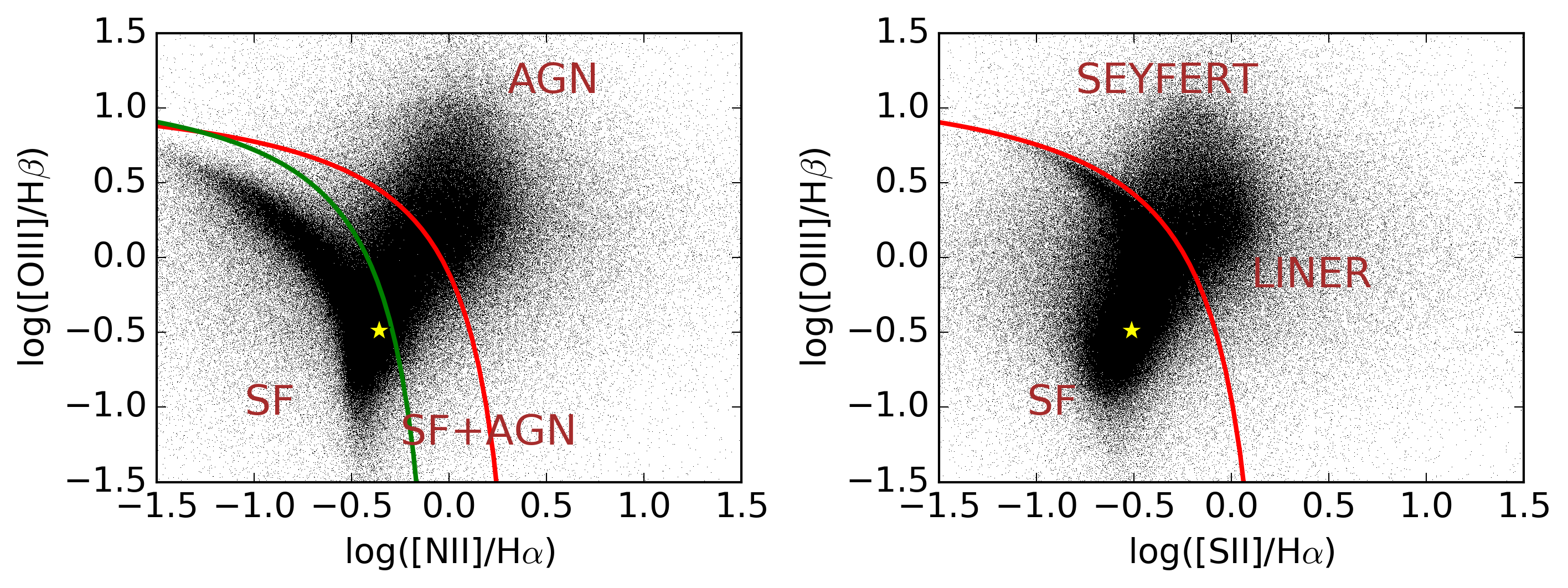}
      \caption{Baldwin-Phillips-Terlevich diagrams according to the \cite{Kewly2006} classification 
scheme, produced using SDSS DR7 data. The yellow star indicates the position of UGSC in the 
diagrams. The error bars are smaller than the symbol size and thus cannot be depicted. The red curve 
(Ke01) denotes the demarcation between star-forming galaxies and AGN as defined by 
\cite{Kewley2001}. The green curve (Ka03) shows the same demarcation as defined by 
\cite{Kauffmann2003}. SF stands for a star-forming galaxy region.}
         \label{fig:bpt}
   \end{figure*}

We identified H$\rm{_\beta}$ 4863\AA{} and [O{\sc III}] 5008\AA{} lines, as well as 
${\rm{H_\alpha}}$ 6565\AA{}, [N{\sc II}] 6550\AA{}, 6585\AA{}, and [S{\sc II}] 6718\AA{}, 6733\AA{} 
lines.  Using these lines, we calculated the redshift of the source: $z=0.0609\pm0.0004$. 
According to the intensity ratios of the emission lines, $\log({{\rm{[N{\sc II}}}] 
_{6585}\over{{\rm{H_\alpha}}}}){=}-0.360{\pm}0.005$,  $\log({{\rm{[S{\sc II}}}]_{6718, 
6733}\over{{\rm{H_\alpha}}}}){=}-0.512{\pm}0.005$, and $\log({{\rm{[O{\sc 
III}}}]\over{{\rm{H_\beta}}}}){=}-0.49{\pm}0.03$, we determined the position of the UGSC on the  
Baldwin-Phillips-Terlevich (BPT) diagrams, as revised by \cite{Kewly2006}. It is shown in 
Fig.~\ref{fig:bpt}. BPT diagrams were originally presented by \cite{Baldwin1981} and are used as 
diagnostic diagrams to classify galaxies based on their emission lines. UGSC lies in the 
star-forming region of the diagram, right below the Ka03 line. In Sect ~\ref{sec:nature} we show 
that UGSC is most likely a starburst galaxy with an AGN core.

\subsection{X-ray data}
\label{sec:xray}

In order to collect additional information on UGSC, we proceeded to acquire X-ray data. The field 
around the source was observed on 9 September 2018 under our ToO request with XRT onboard 
\textit{Neil Gehrels Swift} observatory \citep{Burrows2005}. The total exposure time was 1900 sec. 
 We used the {\tt{xselect}} FTOOL\footnote{\url{http://heasarc.gsfc.nasa.gov/ftools/}} 
\citep{Blackburn1995} with the cleaned event files  produced from the standard pipeline in order to 
extract an image  in the 0.5-8.0 keV band. 
 We do not detect any significant X-ray source at the location of the candidate optical counterpart. 
 Based on the number of counts within an aperture of 0.5\arcsec radius (and an estimate of the 
background from a nearby source-free region), we calculate a source intensity of  
$9.89^{+10.2}_{-6.9}\times10^{-4}$ count/sec (0.5-8.0 keV)  at  68\% confidence based on the BEHR 
algorithm \citep{Park2006}.
 Assuming a $\Gamma=1.7$ power-law model absorbed by the Galactic line-of-sight column density 
($\rm{N_{H}=6.6\times10^{20}}$\,$\rm{cm^{-2}}$ \citep{Dickey1990}),
 this count rate corresponds to an observed flux of $3.85^{+3.97}_{-2.69}\times10^{-14}$ 
$\rm{erg/s/cm^{2}}$ and a luminosity of $3.7^{+3.8}_{-2.6}\times10^{41}$\,$\rm{erg/s}$ (0.5-8.0\,keV 
band). Therefore, we consider $\sim7.4\times10^{41}$\,$\rm{erg/s}$ as the 68\% confidence upper 
limit on its X-ray luminosity.

\subsection{Broadband spectral energy distribution}

\begin{figure}
   \centering
   \includegraphics[width=9cm]{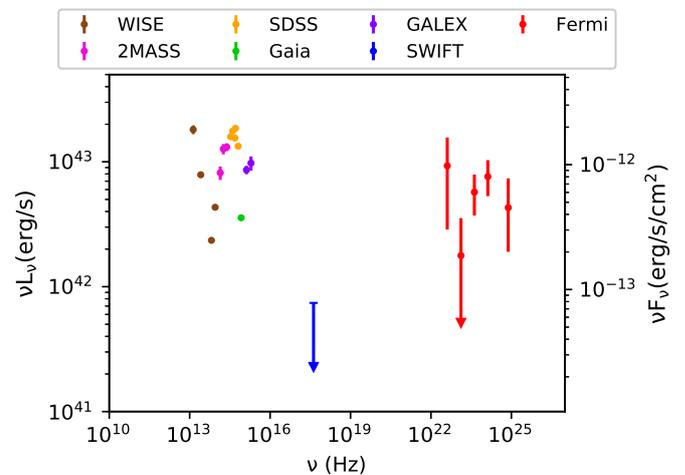}
      \caption{Spectral energy distribution for the UGSC and \g-ray data for 3FGL~J0221.2+2518 from 
3FGL. Red points correspond to \textit{Fermi} data for the UGS, while the others correspond to 
measurements for UGSC.}
         \label{fig:SED}
\end{figure}

We collected archival broadband photometry available for UGSC including measurements by the WISE, 
2MASS, SDSS, Gaia, and GALEX surveys. The combined spectral energy distribution (SED) including the 
X-ray upper limit from the previous section is shown in Fig.~\ref{fig:SED}. It shows signs of three 
peaks that are presumably consistent with a presence of three emission components. The mid-infrared 
peak can be produced by a dusty torus. The peak with a maximum in the optical bands can be caused by 
the underlying galaxy stars emission. While the possible rise towards hard UV in the GALEX bands 
could be explained by the accretion disk emission. However, we note that the data from the surveys  
are not contemporaneous and may represent the source at different activity states, which in turn can 
cause a  semblant multicomponent SED.

We plot the \g-ray data of 3FGL~J0221.2+2518 from 3FGL on the same SED under the assumption that it 
is associated with UGSC. The overall SED shape and its components are consistent with the SEDs of 
known \g-ray emitting NLSy1 \citep{Foschini2012,Paliya2018}. However, the luminosity 
$L_\gamma=3.7\pm0.9\times10^{43}erg/s$ of the UGSC is more than an order of magnitude lower than any 
other NLSy1. Moreover, near- to mid-infrared colors, $[3.4\mu{m}] {-} [4.6\mu{m}] {=} 0.32$~mag and 
$[4.6\mu{m}] {-} [12\mu{m}] {=} 4.24$~mag, are atypical for blazars or NLSy1 \citep[cf.  Fig.~2 from 
][]{Paliya2018}.

\subsection{Nature of UGSC} \label{sec:nature}

Based on the whole set of data, the nature of the UGSC seems to be complex. The information on its 
polarization value may denote the existence of a relativistic jet attributed to an AGN. This 
hypothesis is supported by the shape of the SED and the possible emitting components that are 
typical for low-luminosity AGN. However, the near- to mid-infrared colors of UGSC are not consistent 
with this hypothesis and suggest a different type of source.

The UGSC lies below the Ka03 line in the BPT, but rather close to the composite objects' area. Based 
on the rest of the data presented above, we suggest that there is an AGN contribution to the UGSC 
emission. This argument is also supported by \cite{Kewly2006}. As the authors discuss,   an AGN 
contribution is likely on the condition that $\log({{\rm{[N{\sc II}}}] \over{{\rm{H_\alpha}}}}) 
{\ge} {-0.5}$. Furthermore, based on Fig. 2 in the same work, the UGSC is consistent with the 
positions of an AGN on the [SII]/${\rm{H_\alpha}}$ diagram.

By inspection of the shape of the spectrum, UGSC could be categorized as a type 2 AGN. It is typical 
for this kind of active galaxy to display only narrow lines and a  high degree of polarization, as 
observed, due to scattering from a dusty torus that obscures the nucleus (see, e.g.,  
\citealt{Tadhunter2008}). However, these objects usually have a ratio of $\log({{\rm{[O{\sc 
III}}}]\over{{\rm{H_\beta}}}})\ge{0.48}$ \citep{Shuder1981}.

The upper limit on the X-ray luminosity of UGSC ($\sim7.4\times10^{41}$\,$\rm{erg/s}$) is consistent 
with a classification as a star-forming galaxy, but it cannot rule-out the possibility that it hosts 
a low-luminosity or a heavily obscured AGN. Finally, the optical light curve of the UGSC 
(Fig.~\ref{UFO_lc}) suggests that it can be a variable source, which strengthens the AGN argument.
We conclude that the UGSC is most likely a composite object, i.e., a starburst galaxy with an AGN 
core.

\section{Association with the \textit{Fermi} field and other candidates} \label{sec:candidates}

\begin{figure*}
   \centering
   \includegraphics[width=0.94\textwidth]{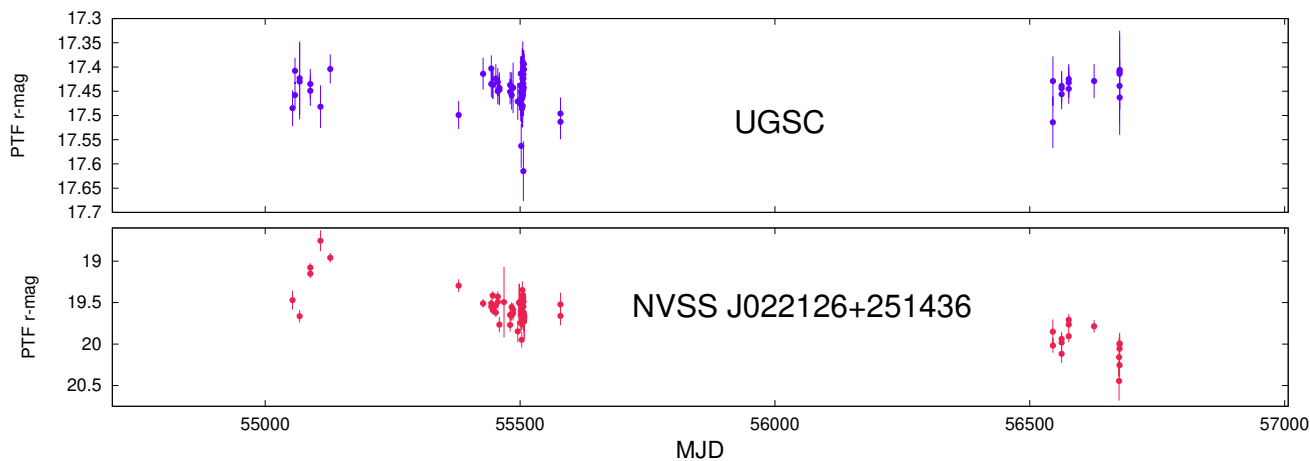}
      \caption{Optical light curves of UGSC and NVSS~J022126+251436 from the Palomar Transient Factory.}
       \label{UFO_lc}
   \end{figure*}

The UGSC could be the potential counterpart for the \g-ray field 3FGL~J0221.2+2518, but a confident 
association is challenging. There are three other candidates proposed for this field. 
\cite{Schinzel2017} introduced two different counterparts. One of them is a known Radio Galaxy, 
NVSS~J022126+251436, 2.3 arcmin away from UGSC, with AGN-like spectral energy distribution and 
variable optical flux (labeled  `RG' in Fig.~\ref{Field}). This source is listed in SDSS DR8 as a 
$r=19.05$ mag source and its polarization degree is consistent with zero due to high measurements 
uncertainties. In Fig.~\ref{UFO_lc} we show the Palomar Transient Factory r-band light curve of this 
source together with the light curve of the UGSC.

The second counterpart by \cite{Schinzel2017} is proposed to be a radio source  located at 
RA=02h21m15.67s, Dec=+25\degr16\arcmin58.48\arcsec. Its position is labeled  `X' in Fig.~\ref{Field} 
and is denoted as a radio source. There is no known source within the 0.3\arcsec\ error radius they 
propose in optical catalogs or in our dataset. The source can be fairly faint and only detectable in 
radio band. Given the lack of information regarding this source, we cannot exclude the possibility 
that it is associated with the 3FGL~J0221.2+2518 field.

Finally, we examined the work by \cite{Massaro2016}, who report that they  have identified a 
counterpart for 3FGL~J0221.2+2518. They collected optical spectroscopic data for the counterpart 
proposed by \cite{Paggi2014} for this field, and classified it as a QSO. The QSO is located at 
RA=02h20m51.24s, Dec=+25\degr09\arcmin27.6\arcsec, which places it at $\sim5\sigma$ uncertainty 
ellipse of the \textit{Fermi} field. It is not shown in Fig.~\ref{Field} since it is relatively far 
from the center of the \g-ray field. Given the location of the QSO with respect to the UGS position 
and its uncertainties, it is extremely unlikely ($\wp=6\times10^{-7}$) that this AGN can be 
associated with the UGS.

\section{Conclusions} \label{sec:concl}

We proposed optical polarimetry as a fast and efficient tool for identifying blazars in a 
high-polarization state as possible counterparts of \g-ray sources from the \textit{Fermi}-LAT 
catalogue. This technique can serve as a powerful addition to a variety of previously proposed 
methods. Moreover, it can be improved by using multiple measurements of a given field at different 
epochs. Thus, variability of polarization (another distinct property of blazars) can be used for 
their identification.

We measured the optical polarization of sources in the 3FGL~J0221.2+2518 field and discovered a new 
extragalactic source positioned at RA=02h21m33.3s, Dec=+25\degr12\arcmin47.3\arcsec, with redshift 
$z {=} 0.0609 {\pm} 0.0004$. Its fractional polarization $5.2\pm1.3\%$ is significantly higher than 
the average polarization of the field $0.91\pm0.07\%$. Analysis of the multiband archival data in 
combination with optical spectroscopy leads us to the conclusion that the source is most likely a 
complex source comprised of an AGN along with a star-forming region in its galaxy.

This result confirms our theoretical estimates, demonstrates the usefulness of our method, and 
motivates its use for future research. The upcoming large polarimetric survey \textit{PASIPHAE} 
\citep{Pasiphae2018} aims to map the polarization of millions of objects in both the northern and 
the southern hemispheres.  \textit{PASIPHAE} will provide an exceptional opportunity to discover 
many previously unknown synchrotron emitters, including dozens of candidate counterparts for 
unidentified \textit{Fermi}-LAT sources.

\begin{acknowledgements}
We thank Richard Britto and Marcello Giroletti for useful comments that improved this work. We thank 
the Swift Observatory Team, in particular  Brad Cenko, for their rapid approval and scheduling of 
ToO observations. The {\em RoboPol} project is a collaboration between Caltech in the USA, MPIfR in 
Germany, 
Toru\'{n} Centre for Astronomy in Poland, the University of Crete/FORTH in Greece, and IUCAA in 
India. D.B. acknowledges support from the European Research Council (ERC) under the European Union’s 
Horizon 2020 research and innovation program under grant agreement No 771282. G.\,V.\,P. 
acknowledges support by the European Commission Seventh Framework Programme (FP7) through the Marie 
Curie Career Integration Grant PCIG-GA-2011-293531 ``SFOnset''. K.K., A.Z., and R.S.  acknowledge 
funding from the European Research Council under the European Union’s Seventh Framework Programme 
(FP/2007-2013)/ERC Grant Agreement n. 617001.
This research has made use of the NASA/IPAC Infrared Science Archive, which is operated by the Jet 
Propulsion Laboratory, California Institute of Technology, under contract with the National 
Aeronautics and Space Administration. This publication makes use of data products from the 
Wide-field Infrared Survey Explorer, which is a joint project of the University of California, Los 
Angeles, and the Jet Propulsion Laboratory/California Institute of Technology, funded by the 
National Aeronautics and Space Administration
This research has made use of the SVO Filter Profile Service 
(http://svo2.cab.inta-csic.es/theory/fps/) supported by the Spanish MINECO through grant 
AyA2014-55216,
The SVO Filter Profile Service (Rodrigo, C., Solano, E., Bayo, A. http://ivoa.net/documents/Notes/SVOFPS/index.html)
and The Filter Profile Service Access Protocol (Rodrigo, C., Solano, E. 
http://ivoa.net/documents/Notes/SVOFPSDAL/oke-index.html).
\end{acknowledgements}

\bibliographystyle{aa}
\bibliography{bibliography}
\end{document}